# Evolution of proto-galaxy-clusters to their present form: theory and observations

Carl H. Gibson [1,2]

[1] University of California San Diego, La Jolla, CA 92093-0411, USA
[2] cgibson@ucsd.edu, http://sdcc3.ucsd.edu/~ir118

and

Rudolph E. Schild[3,4]

[3] Center for Astrophysics, 60 Garden Street, Cambridge, MA 02138, USA
[4] rschild@cfa.harvard.edu

## ABSTRACT

From hydro-gravitational-dynamics theory HGD, gravitational structure formation begins 30,000 years ($10^{12}$ s) after the turbulent big bang by viscous-gravitational fragmentation into super-cluster-voids and $10^{46}$ kg proto-galaxy-super-clusters. Linear and spiral gas-proto-galaxies GPGs are the smallest fragments to emerge from the plasma epoch at decoupling at $10^{13}$ s with Nomura turbulence morphology and length scale $L_N \sim (\gamma\nu/\rho G)^{1/2}$ $\sim 10^{20}$ m, determined by rate-of-strain $\gamma$, photon viscosity $\nu$, and density $\rho$ of the plasma fossilized at $10^{12}$ s. GPGs fragment into $10^{36}$ kg proto-globular-star-cluster PGC clumps of $10^{24}$ kg primordial-fog-particle PFP dark matter planets. All stars form from planet mergers, with ~97% unmerged as galaxy baryonic-dark-matter BDM. The non-baryonic-dark-matter NBDM is so weakly collisional it diffuses to form galaxy cluster halos. It does not guide galaxy formation, contrary to conventional cold-dark-matter hierarchical clustering CDMHC theory ($\Lambda$=0). NBDM has ~97% of the mass of the universe. It binds rotating clusters of galaxies by gravitational forces. The galaxy rotational spin axis matches that for low wavenumber spherical harmonic components of CMB temperature anomalies and extends to $4.5 \times 10^{25}$ m (1.5 Gpc) in quasar polarization vectors, requiring a big bang turbulence origin. GPGs stick together by frictional processes of the frozen gas planets, just as PGCs have been meta-stable for the 13.7 Gyr age of the universe.

## INTRODUCTION

The standard model of cosmology is in the process of rapid decomposition as a relentless flood of new data confronts old ideas (Jeans 1902, Darwin 1889) about fluid mechanics. New space telescopes cover an ever-widening band of frequencies. Ground based tele-scopes are linked and controlled by ever more powerful computers that track events as they happen and freely distribute all information nearly real time on the internet. The standard model of cosmology is cold-dark-matter hierarchical-clustering CDMHC based on the acoustic length scale proposed (Jeans 1902) as the single criterion for gravitational structure formation (see Table 1). In 1902 only the Milky Way nebula of stars was recognized as a galaxy. As pointed out in their book (Hoyle, Burbidge and Narlikar 2000), the spiral nebula (galaxy) Messier 51 had been detected in 1855 by Lord Rosse, and it was only speculated that such objects were Milk-Way-like galaxies, a view strongly dismissed by Agnes Clerke in her 1905 well-known popular book *The System of Stars* based on her perception of the results and conclusions of professional astronomers of the day:

> *The question whether nebulae are external galaxies hardly any longer needs discussion. It has been answered by the progress of research. No competent thinker, with the whole of the available evidence before him, can now, it is safe to say, maintain any single nebula to be a star system of co-ordinate rank with the Milky Way. A practical certainty has been attained that the entire contents, stellar and nebula, of the sphere belong to one mighty aggregation, and stand in ordered mutual relations within the limits of one all embracing scheme (Clerke 1905).*

As Hoyle et al. 2000 note in their preface, pressures of big science funding have badly corrupted the peer review system of astrophysics and cosmology. Papers that deviate from the standard model are dismissed out of hand by referees and scientific editors who depend on big science funds for survival. It was first pointed out theoretically (Gibson 1996) and verified observationally (Schild 1996) that the Jeans 1902 fluid mechanical analysis of standard cosmology is fatally flawed. A new cosmology modified by modern fluid mechanics termed hydro-gravitational-dynamics HGD (Gibson 2009ab) has emerged, but publication of this information has only recently been permitted in a physics journal (Niewenhuizen, Gibson & Schild 2009). Contrary to the inviscid, linear-perturbation-stability analysis of Jeans 1902, gravitational instability is highly non-linear, dominated by viscosity, and absolute (Gibson 1996, 2000). Anti-gravity forces result from negative turbulent and gluon-viscosity stresses (Gibson 2004, 2005), not dark energy ($\Lambda$=0). Viscous and turbulent forces determine all gravitational plasma structure formation. Diffusivity of the nearly-collision-less non-baryonic-dark-matter (neutrino) NBDM (Nieuwenhuizen 2009) prevents Jeans condensation and hierarchical clustering of (mythical) CDM halos during the plasma epoch before decoupling. Artificial "Plummer forces" (Plummer 1911) introduced to fit data from observations by numerical simulations (see Table 1 and Dehnen 2001) compensate for the physical impossibility of CDM halo formation and clustering. The "Plummer force length scales" match the Nomura scale $L_N = 10^{20}$ m and are required to permit numerical simulations to match super-void observations (Tinker and Conroy 2008). The Nomura scale and the observed gas-proto-galaxy-cluster GPG morphology demonstrate effects of weak turbulence at the end of the plasma epoch (Gibson & Schild 2009), as well as the crucial role of PGC clumps of planets as the dominant mass of galaxies (Schild 1996).

Proto-galaxy-clusters form at the last stage of the plasma epoch guided by weak turbulence along vortex lines produced as expanding proto-supercluster-voids encounter fossil density gradients of big bang turbulence, producing baroclinic torques and turbulence on the expanding void boundaries. In this paper we focus on the evolution of proto-galaxy-clusters during the present gas epoch. Examples of gas proto-galaxy-clusters GPGs are shown in Figure 1, from the Hubble Space Telescope Advanced Camera for Surveys HST/ACS observing the Tadpole galaxy merger UGC 10214 (Elmegreen et al. 2004ab). We see that these dimmest objects (magnitude 24-28) with $z > 0.5$ reflect their formation along turbulent vortex lines of the plasma epoch, and clearly reflect the gentle nature of the early universe as these gas-proto-galaxy-clusters GPGs expand ballistically and with the expansion of the universe against frictional forces of the baryonic dark matter (PGC friction). The mechanism of momentum transfer between gas proto-galaxies is much better described by the Darwin 1889 meteoroid collision mechanism (Darwin 1889) than the misleading collisionless-gas mechanisms (Jeans 1902) that permeate ΛCDMHC cosmology and causes it to fail.

From HGD, the chains of star clumps shown in Fig. 1 have been incorrectly identified as “chain galaxies” since their discovery in the Hawaiian Deep Field (Cowlie et al. 1995) at magnitude 25-26. The rows of clumps are not edge-on spiral galaxies (Elmegreen et al. 2004b) and the tadpoles are not end-on chain galaxies (Elmegreen et al. 2004a). Instead, they are GPG chain-clusters of proto-galaxies fragmented along turbulent vortex lines of the primordial plasma (Gibson & Schild 2009). In the following we show end-on proto-galaxy-clusters can best be explained as the “fingers of God” structures observed in the Sloan Digital Sky Survey II and by the Hickson 1993 Compact Group HCG class of galaxy clusters (Hickson 1993) exemplified by Stephan’s Quintet (Gibson & Schild 2007). A complex system of star wakes, globular star cluster wakes and dust trails indicates that the dark matter halo is dominated by PGCs of planets from which the stars and GCs form on agitation, and that the SQ galaxies have separated from a primordial-plasma chain-proto-galaxy-cluster.

The linear gas-proto-galaxy clusters of Fig. 1 show the universe soon after decoupling must have been quite gentle for them to survive. This contrasts with the standard model of galaxy formation where the first galaxies are CDM haloes that have grown by hierarchical clustering to about $10^{36}$ kg (a globular cluster mass) and collected a super-star amount of gas in their gravitational potential wells, about $10^{32}$ kg (100 solar mass). As soon as the gas cools sufficiently so that its Jeans scale permits condensation it does so to produce one super-star and one extremely bright supernova.

The combined effect of these mini-galaxy supernovae is so powerful that the entire universe of gas is re-ionized according to CDMHC. The problem with this scenario is that it never happened. It rules out the formation of old globular star clusters that require very gentle gas motions. The extreme brightness of the first light is not observed. Reionization is not necessary to explain why neutral gas is not observed in quasar spectra once it is understood that the dark matter of galaxies is frozen PFP primordial planets in PGC clumps.

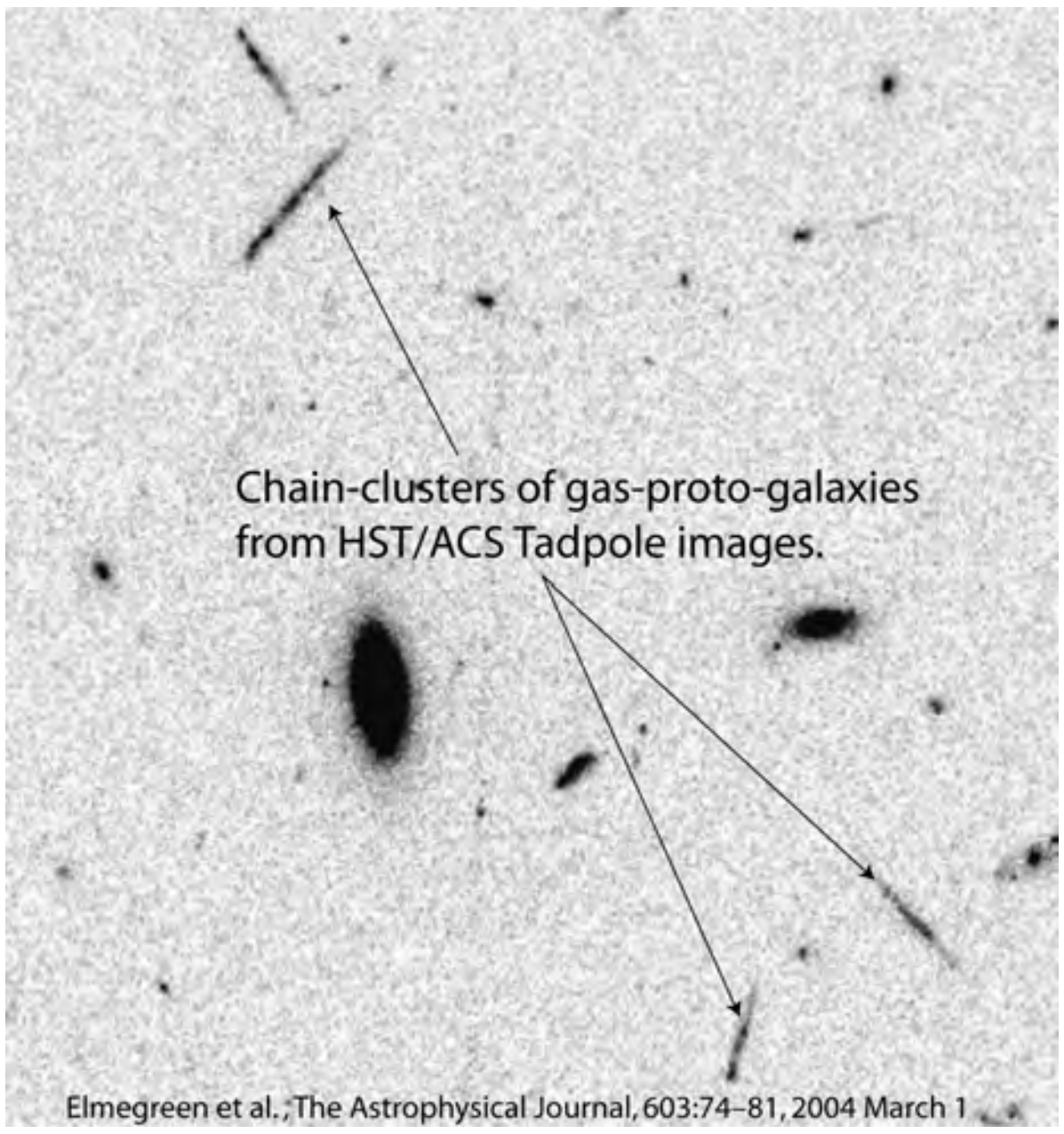

Figure 1. Chain-clusters of gas-proto-galaxies GPGs reflect their origin by gravitational fragmentation along turbulent vortex lines of the plasma epoch at decoupling time t ~ $10^{13}$ s (300,000 years) after the big bang. Only ~0.1% of the baryonic dark matter BDM (planets in clumps) has formed old-globular-star-cluster OGC small stars in these proto-galaxies. More than 80% of the dimmest proto-galaxies (magnitudes 24-28) are in linear proto-galaxy-clusters termed chains, doubles and tadpoles (Elmegreen et al. 2004ab). The tadpole tails and the luminosity between all GPGs are stars formed from frictional BDM planets that form gas when agitated and accrete to form stars.

The chain-clusters of GPGs in Fig. 1 confirm the prediction of HGD that the early gas universe was quite gentle, contrary to CDM where GPGs appear by violent mergers late in the gas epoch at very small scales compared to galaxy sizes observed today. From HGD, each GPG in the linear clusters of several has about $10^{42}$ kg of BDM dark matter

PGCs. The foreground elliptical and spiral galaxies shown in Fig. 1 have not acquired mass by CDMHC merging. The BDM planet-clumps gradually diffuse out of the $10^{20}$ m proto-galaxy core to form the $10^{22}$ m BDM halo against PGC frictional forces. The PGC frictional forces inhibit the ballistic growth of the linear GPGs as well as their growth due to the expansion of the universe. PGCs trigger the formation of star trails and young globular star clusters as they move through the BDM halo.

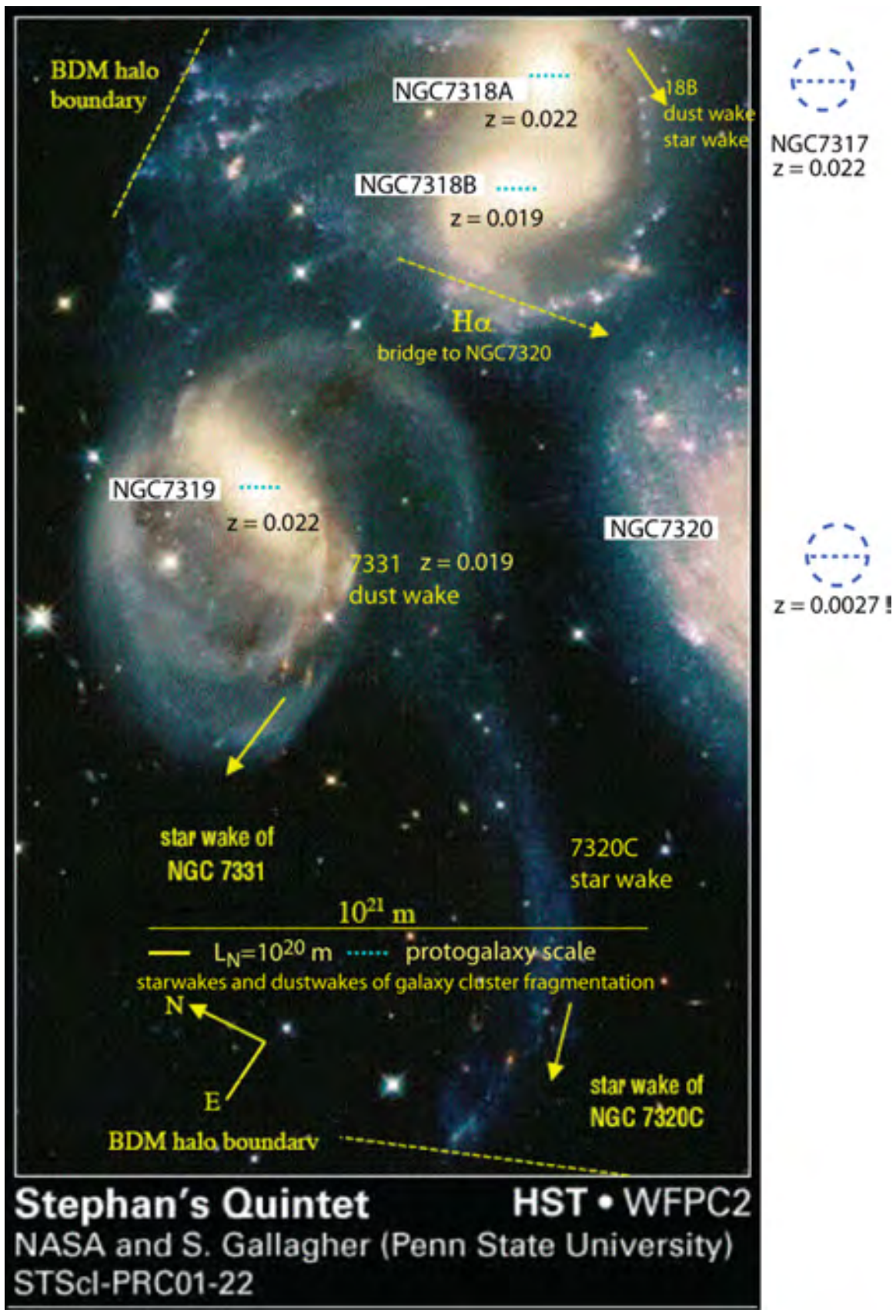

Figure 2. Stephan's Quintet (SQ, HCG 92, Arp 319, VV 288) has long been a very mysterious galaxy cluster (Stephan 1877). The Trio NGC7319, NGC7318A, NGC7317 have redshift 0.022, NGC 7318B has redshift 0.019 that matches that of NGC7331, but nearby NGC7320 has redshift 0.0027! All galaxies are connected by star trails, star cluster trails and dust trails suggesting an end-on chain-cluster of separating GPGs formed at $t<10^{13}$ s along a turbulent vortex line (Gibson & Schild 2002).

An end-on linear GPG example is shown in Figure 2, the spectacular Stephan's Quintet SQ. Three galaxies in a narrow range of angles have precisely the same redshift $z$ = 0.022, one has $z$ = 0.019 but one has a highly anomalous $z$ = 0.0027. SQ (HGC 92) is one of many highly compact galaxy clusters termed Hickson Compact Groups (Hickson 1993). Nearly half of the HCG galaxy clusters have at least one highly anomalous red-shift member. From HGD, gas-proto-galaxies are quite sticky from PGC friction. Even though plasma-proto-galaxies are stretching apart along turbulent vortex lines with the maximum rate of strain of the turbulence the central GPGs formed may not be able to separate. SQ is commonly described as a violent galaxy merger. It is not.

Burbidge and Burbidge 1961 report the highly anomalous nature of the SQ galaxy red-shifts. For virial equilibrium the closely aligned NGC 7313AB galaxies would require an M/L ratio of 300 from dynamical models, but NGC 7320 requires M/L of 10,000, much too large to be credible. A connecting gas bridge to NGC 7320 proves it is not a chance intruder but a separated companion (Gibson & Schild 2007). One possibility claimed (Arp 1973, Arp 1998, Hoyle, Burbidge & Narlikar 2000, Burbidge 2003) is that galaxies may somehow be ejected by an AGN mother galaxy with intrinsic redshifts, which accounts for the observational fact that giant AGN elliptical galaxies are observed with many more nearby quasars than chance will allow. We suggest a more likely possibility is that HCGs like SQ, and various quasar-galaxy associations (Burbidge 2003), are simply end-on views of linear GPGs and where sometimes quasars are included in such end-on linear clusters (see Fig. 8). The Trio is still stuck together by PGC friction and 7313B and 7320 have separated ballistically and from the expansion of the universe. Their close angular proximity is an optical illusion due to their nearness to earth and perspective.

From known properties of the hot big bang universe the Schwarz viscous and turbulent scales of Table 1 show fragmentation will occur early at massive proto-super-cluster scales by formation of expanding super-cluster-voids independent of the NBDM.

Table 1. Length scales of gravitational instability

| Length Scale Name | Definition | Physical Significance |
|---|---|---|
| Jeans Acoustic | $L_J = V_S/(\rho G)^{1/2}$ | Acoustic time matches free fall time |
| Schwarz Viscous | $L_{SV} = (\gamma\nu/\rho G)^{1/2}$ | Viscous forces match gravitational forces |
| Schwarz Turbulent | $L_{ST} = (\varepsilon/[\rho G]^{3/2})^{1/2}$ | Turbulent forces match gravitational forces |
| Schwarz Diffusive | $L_{SD} = (D^2/\rho G)^{1/4}$ | Diffusive speed matches free fall speed |
| Horizon, causal connection | $L_H = ct$ | Range of possible gravitational interaction |
| Plummer force scale | $L_{CDM}$ | Artificial numerical CDM halo sticking length |

$V_S$ is sound speed, $\rho$ is density, G is Newton's constant, $\gamma$ is the rate of strain, $\nu$ is the kinematic viscosity, $\varepsilon$ is the viscous dissipation rate, D is the diffusivity, $c$ is light speed, $t$ is time.

The plan of the present paper is to first review the very different predictions of HGD and CDMHC theories with respect to galaxy formation and evolution. Then Observations are discussed, followed by a Conclusion.

## THEORY

Figure 3 shows the sequence of gravitational structure formation events according to hydro-gravitational-dynamics HGD cosmology leading to primordial gas-proto-galaxies and galaxies of the present time. A hot big bang is assumed at 13.7 Gyr before the present time, followed by an inflation event where big bang turbulent temperature microstructure is fossilized by stretching of space beyond the scale of causal connection $L_H = ct$, where $c$ is the speed of light and $t$ is the time. Gravitational instability produces the first structure in the plasma epoch by fragmentation, as proto-supercluster-voids begin to grow at $10^{12}$ seconds (30,000 years) leaving proto-superclusters in between. The proto-super-clusters do not collapse by gravity but expand with the expansion of space working against the photon viscosity. Viscous dissipation rates can be estimated from $\varepsilon \sim \nu\gamma^2$, giving $\varepsilon \sim 400$ $m^2\ s^{-1}$. Photon-electron collision lengths were $\sim 10^{18}$ m, less than the horizon scale $L_H = 3x10^{20}$ m as required by continuum mechanics. Viscous dissipation rates in the gas epoch after decoupling decreased with $\nu$ and $\gamma$ to $\varepsilon \sim 10^{-13}\ m^2\ s^{-1}$ values small enough to permit formation of dark matter planets in clumps and the small stars of old globular star clusters. First life needs chemicals produced by the death of the first stars.

The criterion for fragmentation is that the Schwarz viscous scale $L_{SV}$ matches $L_H$ (see Table 1). The NBDM decouples from the plasma because it is weakly collisional. The voids grow as rarefaction waves that approach the sound speed $c/3^{1/2}$. Turbulence is produced at expanding void boundaries by baroclinic torques. Observations confirm that the Reynolds number of the turbulence is rather weak. Fragmentations and void formations occur at smaller and smaller scales until the plasma to gas transition (decoupling) at $10^{13}$ seconds (300,000 years). The weak turbulence produces plasma-proto-galaxies by fragmentation, with NBDM filled voids formed along stretching and spinning turbulent vortex lines.

In Fig. 3, a. Cosmic Microwave Background temperature anisotropies reflect structures formed in the plasma epoch. b. From HGD the photon viscosity of the plasma epoch prevents turbulence until the viscous Schwarz scale $L_{SV}$ becomes less than the Hubble scale (horizon scale, scale of causal connection) $L_H = ct$, where $c$ is the speed of light and $t$ is the time. The first plasma structures were proto-super-cluster voids and proto-super-clusters at $10^{12}$ seconds (30,000 years). c. Looking back in space is looking back in time. Proto-galaxies were the last fragmentations of the plasma (orange circles with green halos) at $10^{13}$ seconds. d. The scale of the gravitational structure epoch is only $3 \times 10^{21}$ m compared to present supercluster sizes of $10^{24}$ m and the largest observed super-void scales of $10^{25}$ m. e. Turbulence in the plasma epoch is generated by baroclinic torques on the boundaries of the expanding super-voids.

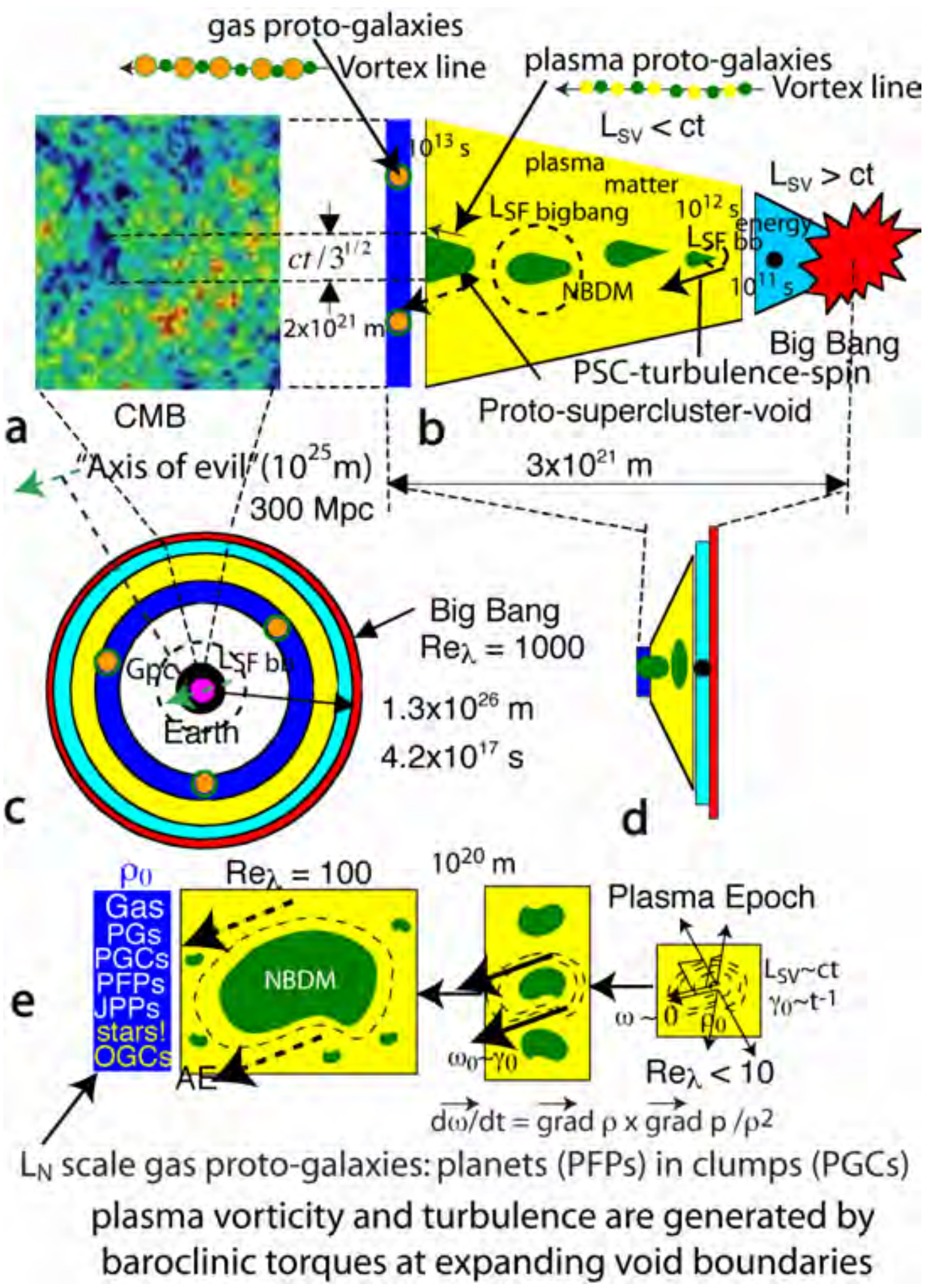

Figure 3. Protogalaxy formation at the end of the plasma epoch by hydro-gravitational-dynamics HGD theory.

Gas chains of proto-galaxies GPGs are formed at decoupling, as shown by the cartoon at the top of Fig. 3a. Because photons suddenly decouple from electrons the viscosity of the fluid suddenly decreases by a factor ~ $10^{13}$, greatly decreasing the viscous Schwarz scale and the fragmentation mass. Two fragmentation scales work simultaneously with the same gravitational free fall time to produce Jeans mass clumps PGCs of earth mass gas planets PFPs, which today is the baryonic dark matter of galaxies.

Figure 4 illustrates super-void and galaxy formation following the standard cosmological model (ΛCDMHC) advocated in the Peebles 1993 book *Principles of Physical Cosmology*. According to the Peebles 1993 timetable (Table 25.1, p 611) (Peebles 1993) superclusters, walls and voids (gaps) form at redshift $z \sim 1$; that is, at least 5 Gyr after the big bang, versus 0.0003 Gyr (Gibson 1996) and HGD voids that are completely empty.

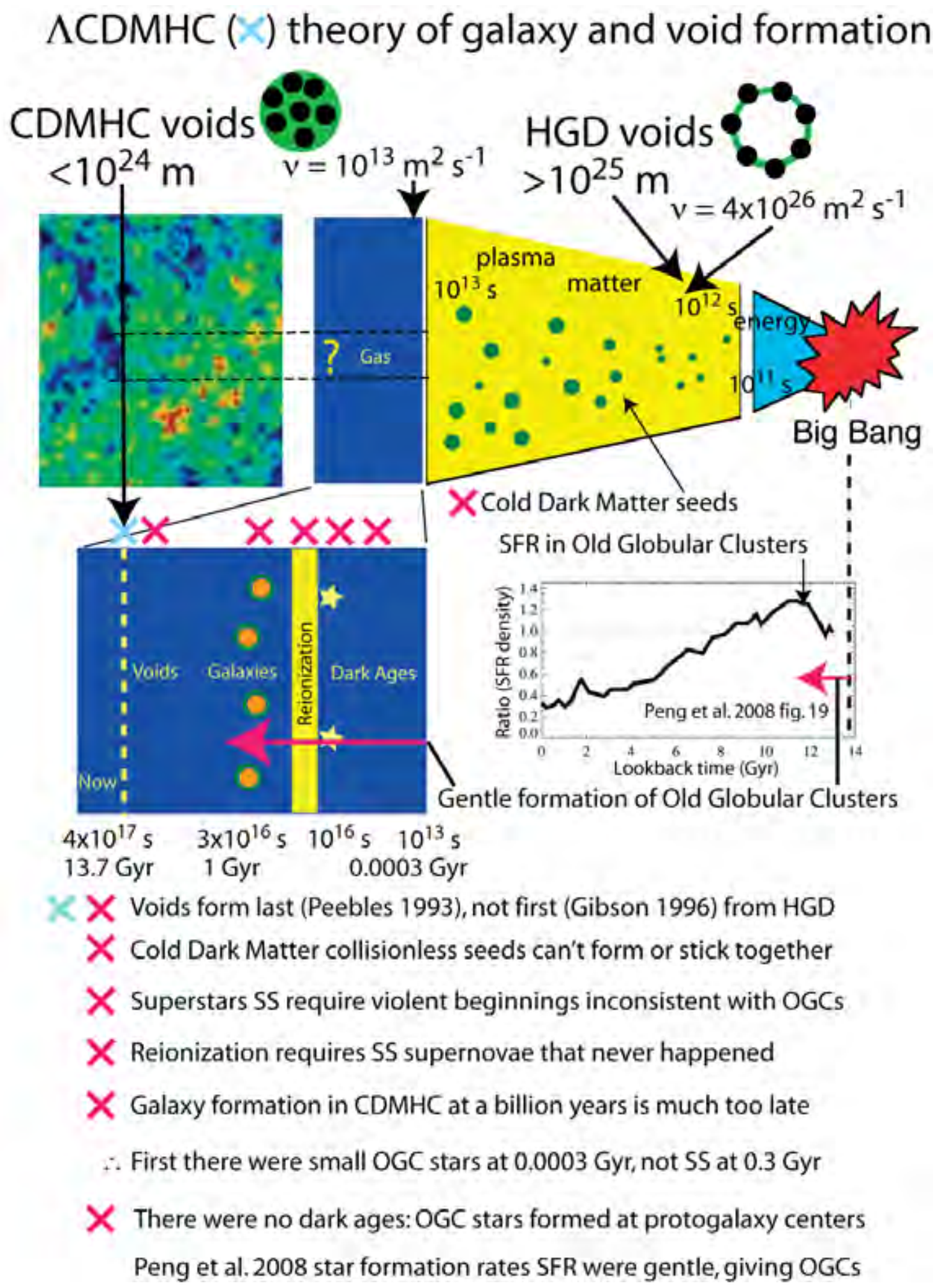

Figure 4. Void and galaxy formation by the standard ΛCDM theory fail (blue Xs) to reconcile with observations and fluid mechanical HGD theory (see text). Other failed aspects of ΛCDMHC theory are indicated by red Xs.

Completely empty super-voids have been detected by radio telescopes with void sizes at least 300 Mpc or $10^{25}$ m, 10% of the horizon scale $L_H = 10^{26}$ m (Rudnick et al. 2008). (Peebles 2007) notes that observations of empty voids on locally observed scales $>10^{24}$ m falsifies any CDMHC model. Black circles indicate $10^{24}$ m superclusters, green is NBDM, white is empty in the void cartoons at the top of Fig. 4.

Fig. 4 contrasts predictions of ΛCDMHC theory with observations and the predictions of HGD theory. HGD cosmology is driven by turbulent combustion at Planck scales from Planck-Kerr instability, with Taylor microscale Reynolds number $Re_\lambda \sim 1000$. Gluon viscosity terminates the event after cooling from $10^{32}$ K Planck temperatures to $10^{28}$ K strong force freeze-out temperatures where quarks and gluons can appear. Turbulent temperature patterns are frozen as turbulence fossils by exponential inflation of space driven by negative stresses of both turbulence and gluon viscosity pulling $\sim 10^{93}$ kg of mass-energy out of the vacuum against the Planck tension $c^4/G$. The mass-energy of our present horizon is only $\sim 10^{53}$ kg, $\sim 10^{-40}$ fraction of the universe produced by the big bang. A black dot in the blue inflation triangle of Fig. 3 symbolizes ~ 20% temperature fluctuations expected from big bang turbulence, contrasting with tiny quantum-mechanical fluctuations expected in the standard model in Fig. 4.

In CDMHC models voids form last rather than first, so this difference is most easily tested by observations. The time of first void formation from HGD is $10^{12}$ s, compared to $\sim 10^{17}$ s for CDMHC. Star formation rates (Peng et al. 2008) favor small old globular-cluster stars. These could not possibly be formed under the violent conditions of galaxy formation and mergers intrinsic to CDMHC. Formation and wide distribution of life also requires the gentle dense clumps and clusters of warm and multiple planets provided by HGD (Gibson & Wickramasinghe 2010, Gibson, Schild & Wickramasinghe 2010).

We now examine available observations for comparison with theories of galaxy formation and evolution.

## OBSERVATIONS

Evidence of the large primordial super-voids of HGD theory (Fig. 1) is shown in Figure 5 (Rudick et al. 2008). Focusing on the direction of the anomalous "cold spot" of the CMB it was found that a $10^{25}$ m (300 Mpc) completely empty region could explain the $\sim 7\times10^{-5}$ $^{o}$K CMB cold spot by the integrated Sachs-Wolfe method. The empty region is estimated to be at redshifts $z \sim 1$, and is therefore completely impossible to explain by CDMHC models where super-voids are formed last rather than first (Fig. 4). The probability of such a void forming from concordance CDM models is estimated (Rudick et al. 2008) to be $< 10^{-10}$.

(Tinker and Conroy 2008) explain the void phenomenon by numerical simulations and produce a relatively empty super-void of scale $10^{24}$ m using numerical simulations and numerically convenient but entirely imaginary "Plummer forces" (Dehnen 2001). As mentioned previously, Plummer forces are physically untenable for weakly collisional NBDM materials such as CDM.

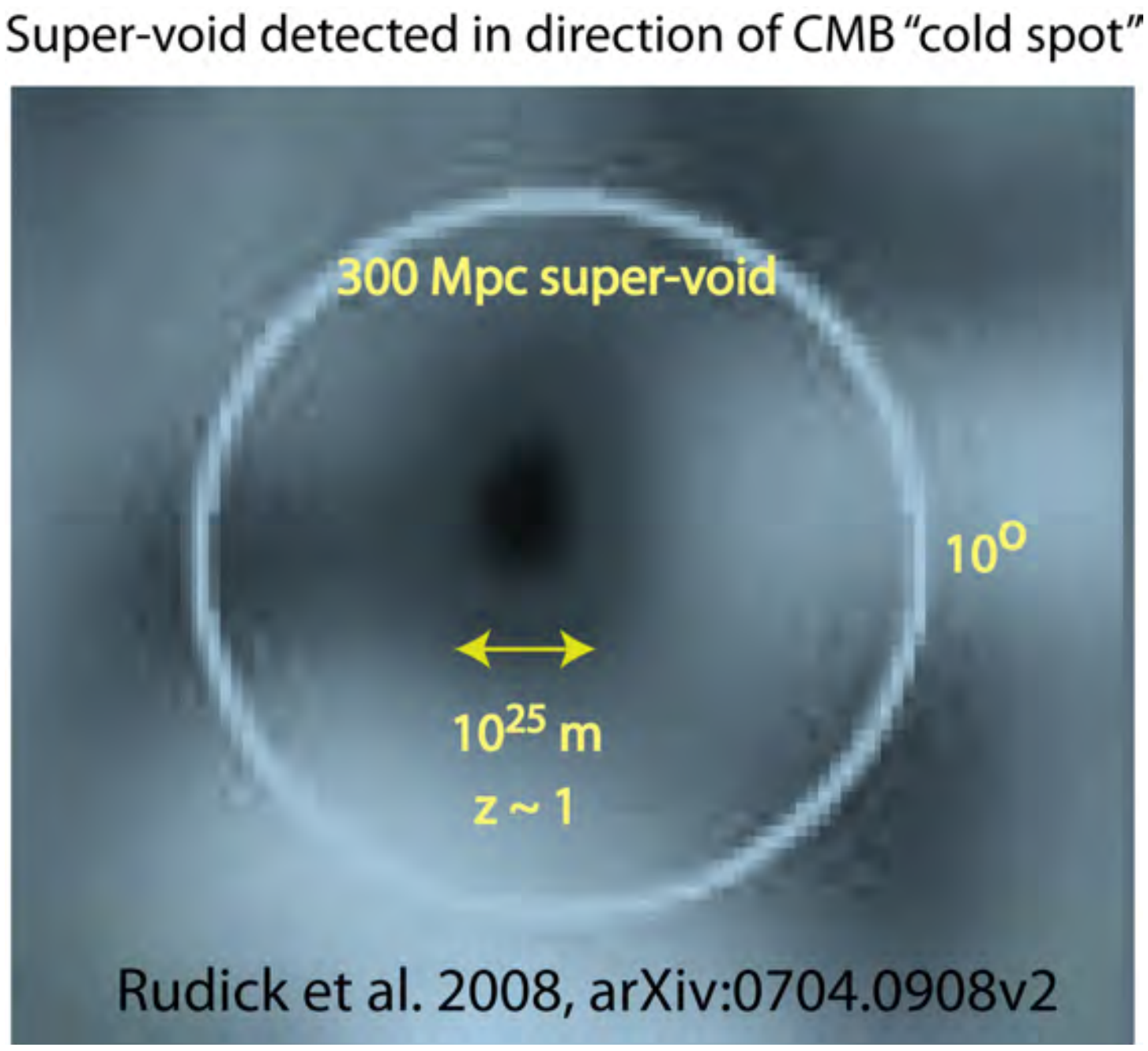

Figure 5. Super-void detection by the radio telescope very large aperture (NVSS) survey in the direction of the anomalous cold spot of the CMB. Such large voids are expected from HGD but are impossible to using CDM models (Rudnick et al. 2008).

Figure 6 summarizes observational evidence that the dimming of supernova Ia events is not evidence for dark energy and a cosmological constant Λ, but is merely a systematic error due to the presence in the near vicinity of shrinking white dwarf stars approaching the Chandrasekhar instability limit of BDM frozen planets partially evaporated by strong spin radiation (1.44 $M_{Sun}$). Dimness of the SNe Ia events increases with their magnitudes, and cannot be explained by uniform grey dust (*ie*: dust without the properties of PFP planets) as shown by the top curve (Reiss et al. 2004).

In Fig. 6, open circles emphasize SNe Ia events unobscured by evaporated BDM planet atmospheres (no dark energy) for the Reiss et al. dimness models. Solid red ovals emphasize events partially obscured by planet atmospheres (non-linear grey dust). Thousands of BDM planets (right insert) in the Helix planetary nebula are evaporated by spin powered radiation from the central white dwarf. From HGD the Helix PNe is not ejected from a massive precursor; instead, the BDM planets are evaporated in place. The observed dimness is caused by fossil turbulence electron density fluctuations in gas with density ρ ~ $10^{-12}$ kg $m^{-3}$ sufficient to permit turbulence. The large 20-30% dimness at z ~ 0.5 cannot be explained by reasonable quantities of dust or gases alone in the observed $10^{13}$ m planet atmospheres shown in the Helix PNe BDM planets insert: it requires fossil electron-density turbulence forward scattering (Gibson et al. 2007).

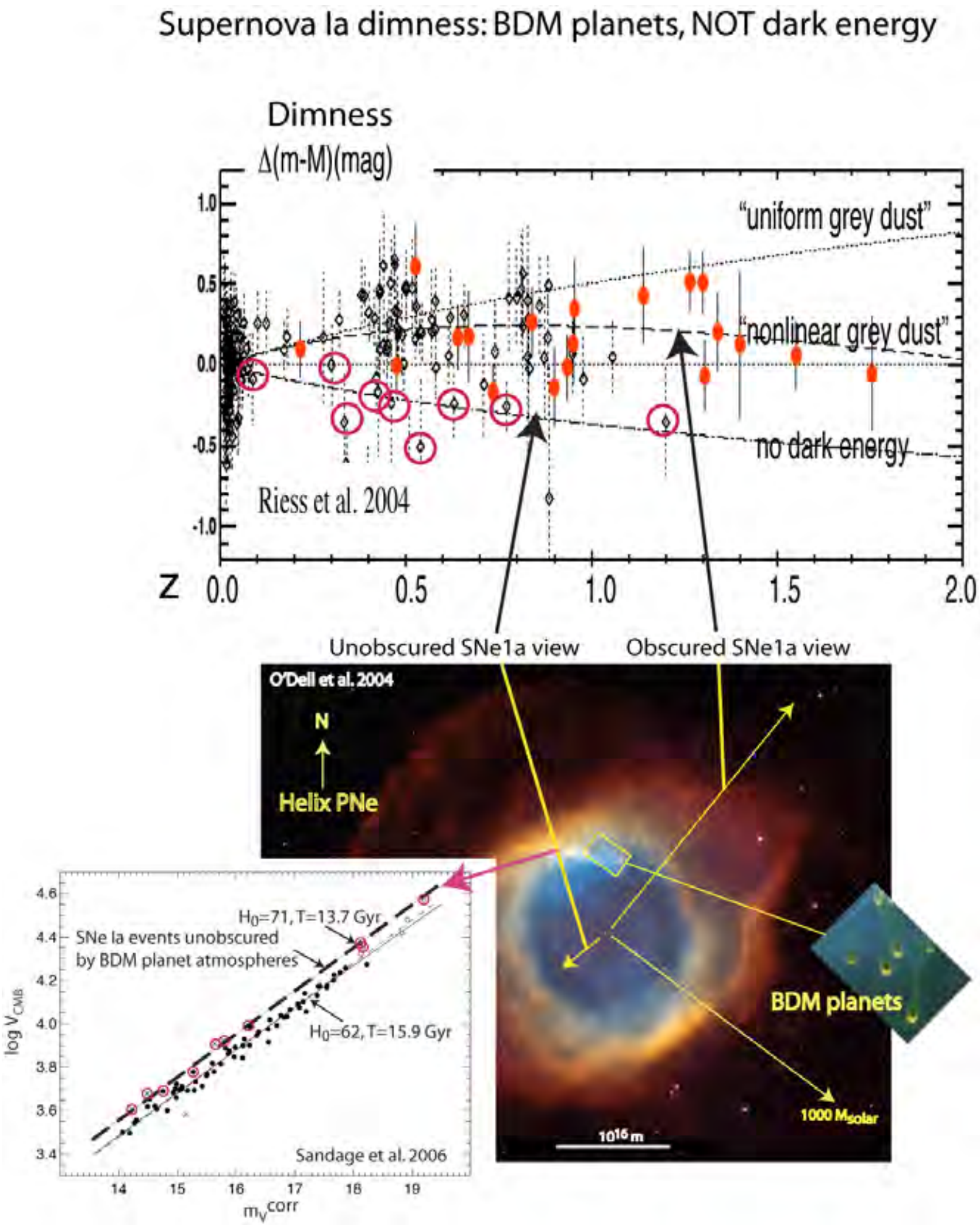

Figure 6. Observations that show the anomalous dimness of supernova Ia events (Reiss et al. 2004) and anomalously low Hubble constants (Sandage et al. 2006) can be attributed to BDM planet atmospheres, not dark energy (Gibson & Schild 2007). The nearby Helix planetary nebula PNe at $6x10^{18}$ m has a central white dwarf with polar jet that evaporates ambient BDM planets of its PGC. A close-up view is shown in the insert on the right (O'Dell 2004).

From HGD, all stars are formed from BDM planets in PGCs. All PGCs have the primordial density $\rho_0 \sim 4x10^{-17}$ kg $m^{-3}$, which matches the density of globular star clusters. The size of the planets, their atmospheres, and separations observed are consistent with this primordial density (Gibson & Schild 2007).

The Fig. 6 insert at lower left shows the Sandage et al. 2006 SNe Ia study of the Hubble Constant $H_0$, carefully corrected for Cepheid variable distances and locations. The age of the universe is 15.9 Gyr from this study, which is unacceptably large. However, open red circles show SNe Ia event lines of sight unobscured by BDM planet atmospheres from HGD. These agree very well with the CMB age of the universe of 13.7 Gyr. Gamma ray burst dimmnesses clinch this interpretation (Gibson & Schild 2009).

Figure 7 shows velocities $V_{LG}$ in km $s^{-1}$ of the local group of galaxies as a function of their distances in Mpc so the slope of a line from the origin is a measure of the Hubble Constant. See references in Gibson & Schild 2009.

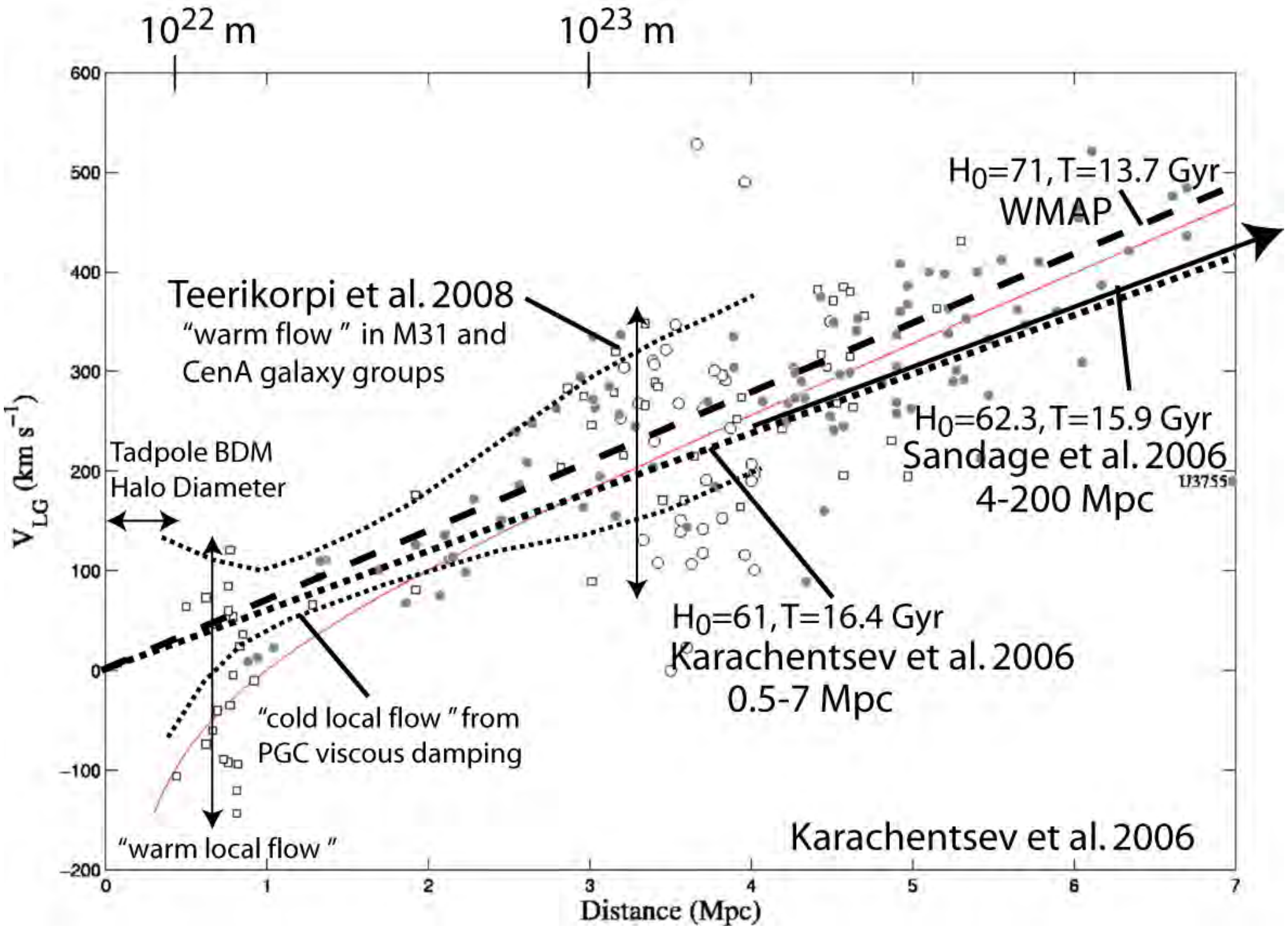

Figure 7. Estimates of the Hubble Constant for galaxies in the local group show wide scatter out to distances of a Mpc due to frictional interactions of BDM halos. The Tadpole BDM halo size is shown by the horizontal double arrow (Gibson & Schild

2002). Beyond this distance the galaxies begin to separate due to Hubble flow. Warm flows in M31 and CenA galaxy groups have $H_0$ dispersions similar to that of galaxies near the Milky Way, as shown by vertical double arrows. The dotted line is an extrapolation to the Sandage et al. 2006 Hubble constants shown in Fig. 6, at 4 to 200 Mpc.

Figure 8 shows a Sloan Digital Sky Survey map of local galaxies compared to the HGD interpretation of Stephan's Quintet as an end-on chain of gas proto-galaxies. In the top panel, note that galaxies with old stars, indicated by red dots, are often aligned in thin pencils termed "fingers of god". Blue dots denote younger galaxies with more blue stars. The reason for this is that the galaxies are relatively near to earth, with redshift z ~ 0.1 or less, so perspective causes a decrease in angular separation for distant galaxies that are already nearly aligned. An arrow shows $10^{25}$ m, about 10% of the present horizon $L_H$. The red pencil-like features are interpreted from HGD as chain clusters of old galaxies aligned by vortex lines in the plasma epoch that have continued moving along these directions ballistically and from the homogeneous straining of the universe during the gas epoch. Dashed circles indicate $10^{24}$ m. PGC frictional stickiness has inhibited separation of the galaxies along their axes and in transverse direction, as shown for Stephan's Quintet.

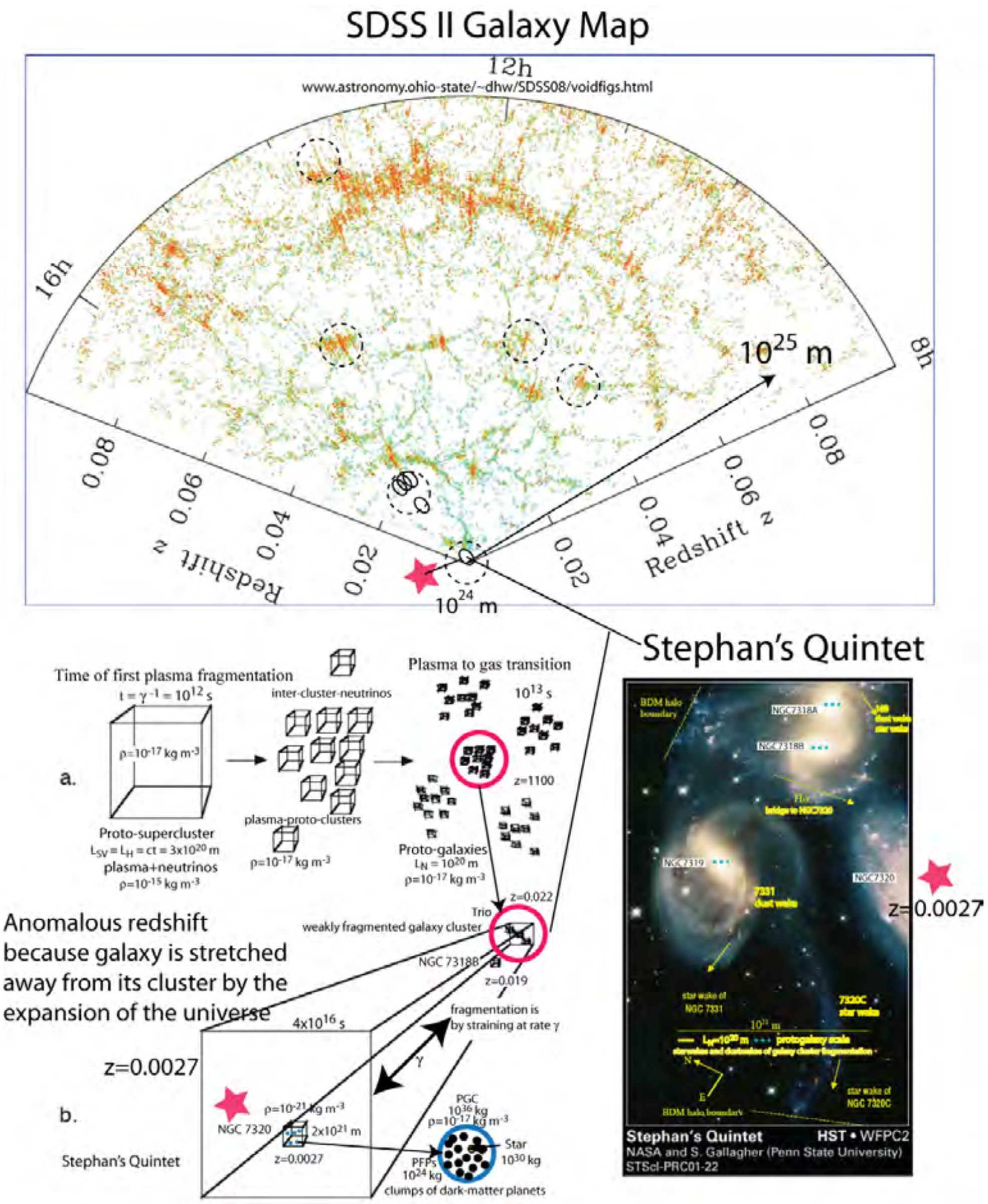

Figure 8. Stephan's Quintet provides evidence that linear galaxy clusters were formed by fragmentation in the plasma epoch along turbulent vortex lines [22]. Red stars indicate the most anomalous galaxy NGC 7320 with redshift $z$ = 0.0027 in three views.

Fig. 8a summarizes the history of SQ formation starting from the time of first fragmentation to plasma to gas transition. Fig. 8b shows SQ at present, with the Trio about $4x10^{24}$ m distant and NGC 7320 with redshift 0.0027 at about $10^{23}$ m. A cartoon of the SQ galaxies is shown near the origin of the SDSS II Galaxy Map.

It seems clear from Fig. 8 that the Trio of SQ galaxies are not clustered by chance or by merging but were formed simultaneously in a linear cluster of proto-galaxies along a turbulent vortex line of the plasma epoch. For 13.7 Gyr they have resisted separation by the expansion of the universe due to PGC friction from frictional interactions of galaxy BDM dark matter planets in galaxy dark matter halos (Gibson & Schild 2007b).

The HGD interpretation of Fig. 8 is that PGC friction inhibits the separation of galaxies in the gas epoch by collisional and tidal interactions of BDM planet halos. The Arp 1973 suggestion that NGC 7331 has ejected the other galaxies with intrinsic redshifts is unnecessary, and would require introduction of an unknown class of new physical laws.

## CONCLUSIONS

Observations exclude the ΛCDMHC standard model for galaxy and void formation as the last steps of gravitational structure formation rather than the first from HGD where the Jeans 1902 theory that provides the basis of CDM is obsolete and fluid mechanically untenable. HGD explains the formation of galaxies as the last stage of gravitational fragmentation starting early in the plasma epoch with proto-super-clusters and proto-super-voids, and finishing with plasma-proto-galaxy morphology determined by weak turbulence from gravitational void expansions and fossil turbulence density gradients from the epoch of strong big bang turbulence. The evolution of gas proto-galaxies from HGD is extremely gentle compared to an unnecessarily violent epoch of super-star formation, supernovae and re-ionization required by ΛCDMHC. These CDM events never happened.

Dimming of SNe Ia events by evaporated BDM planet atmospheres provides an HGD alternative to the new physical laws required by the dark energy hypothesis, Λ, and the Sandage et al. 2006 evidence that the universe age is 15.9 Gyr. The alternative is that there is no dark energy, there is no Λ, and corrections for dimming give a universe age of 13.7 Gyr. Anti-gravity negative stresses needed to produce space-time-mass-energy during the big bang are supplied first by turbulence inertial-vortex forces, and then by gluon-viscous negative stresses during inflation (Gibson & Schild 2010).

Stephan's Quintet confirms predictions of HGD about the evolution of chain gas-proto-galaxy clusters and the importance of PGC friction to stick proto-galaxies together and resist ballistic forces and universe-space-expansion that try to move them apart. The interpretation of SQ and chain-galaxy-clusters by HGD theory provides an alternative to suggestions (Arp 1973, Arp 1998, Hoyle, Burbidge & Narlikar 2000, Burbidge 2003) that central galaxies in chain clusters can emit galaxies and quasars with intrinsic redshifts. Globular cluster wakes, star wakes and dust wakes clearly show the galaxies of SQ were formed in a linear chain and have all separated, never merged, as the galaxy cluster has evolved against the PGC friction of the galaxy-dark-matter-halos consisting of frozen planets in GC-mass clumps.

Further evidence of proto-globular-star-cluster friction from dark-matter-planet interactions is provided by the Hubble diagrams of Fig. 7 for the local group and Fig. 6 from Sandage et al. 2006 at larger distances up to 200 Mpc from SN Ia. PGC-friction explains the random scatter of galaxy velocities in clusters for ~ Mpc lengths scales small enough for BDM halos to interact. At larger scales the expansion of the universe becomes the dominant mechanism to separate galaxies.

## REFERENCES

Clerke, Agnes M. 1905. The System of the Stars, Adam & Charles Black, London UK.

Cowie, L., Hu, E., & Songaila, A. 1995. AJ, 110, 1576.

Darwin, G. H. 1889. On the mechanical conditions of a swarm of meteorites, and on theories of cosmogony, Phil. Trans., 180, 1-69.

Dehnen, W. 2001. Toward optimum softening in 3D N-body codes: Minimizing the force error, Mon. Not. Roy. Astron. Soc., 324, 273-291.

Elmegreen, D. M., Elmegreen, B. G. and Sheets, C. M. 2004a. Chain galaxies in the Tadpole Advanced Camera for Surveys field, ApJ, 603, 74-81.

Elmegreen, D. M., Elmegreen, B. G. and Hirst, C. M. 2004b. Discovery of face-on counterparts of chain galaxies in the Tadpole Advanced Camera for Surveys field, The Astrophysical Journal, 604:L21–L23.

Gibson, C. H. 1968a. Fine structure of scalar fields mixed by turbulence: I. Zero-gradient points and minimal gradient surfaces, Phys. Fluids, 11: 11, 2305-2315.

Gibson, C. H. 1968b. Fine structure of scalar fields mixed by turbulence: II. Spectral theory, Phys. Fluids, 11: 11, 2316-2327.

Gibson, C. H. 1981. Buoyancy effects in turbulent mixing: Sampling turbulence in the stratified ocean, AIAA J., 19, 1394.

Gibson, C. H. 1986. Internal waves, fossil turbulence, and composite ocean microstructure spectra," J. Fluid Mech. 168, 89-117.

Gibson, C.H. 1991. Kolmogorov similarity hypotheses for scalar fields: sampling intermittent turbulent mixing in the ocean and galaxy, Proc. Roy. Soc. Lond. A, 434, 149-164.

Gibson, C.H. 1996. Turbulence in the ocean, atmosphere, galaxy and universe, Appl. Mech. Rev., 49, no. 5, 299–315.

Gibson, C. H. 1999. Fossil turbulence revisited, J. of Mar. Syst., 21(1-4), 147-167, astro-ph/9904237

Gibson, C.H. 2000. Turbulent mixing, diffusion and gravity in the formation of cosmological structures: The fluid mechanics of dark matter, J. Fluids Eng., 122, 830–835.

Gibson, C.H. 2004. The first turbulence and the first fossil turbulence, Flow, Turbulence and Combustion, 72, 161–179.

Gibson, C.H. 2005. The first turbulent combustion, Combust. Sci. and Tech., 177: 1049–1071, arXiv:astro-ph/0501416.

Gibson, C. H. 2006a. Turbulence, update of article in Encyclopedia of Physics, R. G. Lerner and G. L. Trigg, Eds., Addison-Wesley Publishing Co., Inc., pp.1310-1314.

Gibson, C.H. 2006b. The fluid mechanics of gravitational structure formation, astro-ph/0610628.

Gibson, C.H. 2008. Cold dark matter cosmology conflicts with fluid mechanics and observations, J. Applied Fluid Mech., Vol. 1, No. 2, pp 1-8, 2008, arXiv:astro-ph/0606073.

Gibson, C.H., Bondur, V.G., Keeler, R.N., Leung, P.T., Prandke, H., & Vithanage, D. 2007. Submerged turbulence detection with optical satellites, Proc. of SPIE, Coastal Remote Sensing, Aug. 26-27, edited by R. J. Frouin, Z. Lee, Vol. 6680, 6680X1-8. doi: 10.1117/12.732257

Gibson, C.H. & Schild, R.E. 2002. Interpretation of the Tadpole VV29 Merging Galaxy System using Hydro-Gravitational Theory, arXiv:astro-ph/0210583.

Gibson, C.H. & Schild, R.E. 2007a. Interpretation of the Helix Planetary Nebula using Hydro-Gravitational-Dynamics: Planets and Dark Energy, arXiv:astro-ph/0701474.

Gibson, C.H. & Schild, R.E. 2007b. Interpretation of the Stephan Quintet Galaxy Cluster using Hydro-Gravitational-Dynamics: Viscosity and Fragmentation, arXiv[astro-ph]:0710.5449.

Gibson, C. H. and Schild, R. E. 2009. Hydro-Gravitational-Dynamics of planets and dark energy, J. Appl. Fluid Mech., 2(1), 1, arXiv:0808.3228v1.

Gibson, C. H. & Schild, R. E. 2010. Turbulent formation of protogalaxies at the end of the plasma epoch: theory and observations, Journal of Cosmology, 6, 1351-1364.

Gibson, C. H. & Wickramasinghe, N. C. (2010). The imperatives of Cosmic Biology, Journal of Cosmology, in press, arXiv:1003.0091.

Gibson, C. H., Schild, R. E., and Wickramasinghe, N. C. 2010. The Origin of Life from Primordial Planets, In progress, arXiv:1004.0504.

Hickson, P. 1993. Atlas of Compact Groups of Galaxies, Gordon & Breach, New York, New York.

Hoyle, F., Burbidge, G. and Narlikar, J. V. 2000. A Different Approach to Cosmology, From a static universe through the big bang towards reality, Cambridge Univ. Press, Cambridge UK.

Jeans, J. H. 1902. The stability of spherical nebula, Phil. Trans., 199A, 0-49.

Nieuwenhuizen, Th. M. 2009. Do non-relativistic neutrinos constitute the dark matter?, EPL (Europhysics Letters), 86, 59001 (6pp), doi: 10.1209/0295-5075/86/59001.

Nieuwenhuizen, Th. M., Gibson, C. H. and Schild, R. E. 2009. Gravitational hydrodynamics of large-scale structure formation, EPL (Europhysics Letters), 88, 49001 (6pp), doi: 10.1209/0295-5075/88/49001.

O'Dell, C. R., McCullough, P. R. and Meixner, M. 2004. Unraveling the Helix Nebula: Its structure and knots, The Astronomical Journal, 128:2339–2356.

Peebles, P. J. E. 1993. Principles of Physical Cosmology, Princeton University Press, Princeton, NJ.

Peebles, P. J. E. 2007. Galaxies as a cosmological test, arXiv: 0712.2757v1.

Peng, E. W. et al. 2008. The ACS Virgo Cluster Survey. XV. The Formation Efficiencies of Globular Clusters in Early-Type Galaxies: The Effects of Mass and Environment, ApJ, 681, 197-224.

Plummer, H. C. 1911. Mon. Not. Roy. Astron. Soc., 71, 460.

Reiss et al. 2004. Type Ia supernova discoveries at z > 1 from the Hubble Space Telescope: Evidence for past deceleration and constraints on dark energy evolution, ApJ, 607, 665-687.

Rudnick, L., Brown, S. and Williams, L. R. 2008. Extragalactic radio sources and the WMAP Cold Spot, arXiv:0704.0908v2.

Sandage et al. 2006. The Hubble constant: A summary of the HST Program for the luminosity calibration of Type Ia supernovae by means of Cepheids, ApJ, 653, 843.

Schild, R.E & Gibson, C.H. 2008. Lessons from the Axis of Evil, axXiv:astro-ph/0802.3229v2.

Tinker, J. L. and Conroy, C. 2008. The Void Phenomenon Explained, arXiv:0804.2475v2.